\documentclass{aastex}
\begin{document}
\title{RADIAL ARC STATISTICS: \\ 
A NEW POWERFUL PROBE OF THE CENTRAL DENSITY PROFILE OF GALAXY CLUSTERS}
\author{Kohji Molikawa, Makoto Hattori}
\affil{Astronomical Institute, Graduate School of Science, Tohoku University,\\
 Aramaki aza Aoba, Aoba-ku, Sendai, Miyagi, 980-8578, Japan}
\email{molikawa@astr.tohoku.ac.jp, hattori@astr.tohoku.ac.jp}
\begin{abstract}
We show that an expected number ratio of 
radial arcs 
(gravitationally lensed images 
whose major axes lie in the radial direction 
of a cluster-lens potential) 
to tangential arcs 
(gravitationally lensed images 
whose major axes lie in the tangential direction)
has strong dependence 
on the central density profile of galaxy clusters
and has little dependence on other parameters, e.g. cluster temperature, 
background source galaxy redshift etc..
A comparison of the expected number ratios
with observed ratios provides a robust test
to constrain the central density profile of galaxy clusters.
A tentative comparison with the observational data shows
that the central density profile of galaxy clusters is 
$\rho(r) \propto r^{-1 \sim -1.5}$.
This result indicates that the dark matter is collisionless
at least on cluster scale.
Our result gives an upper limit on a collision cross-section 
of self-interaction of dark matter 
($\sigma_{\rm coll}$) as 
$\sigma_{\rm coll}/m<0.1\;{\rm cm^2/g}$ where $m$ is a dark matter particle mass.
\end{abstract}
\keywords{galaxies: clusters: general --- dark matter --- gravitational lensing}
\section{Introduction}
The central density profile of dark matter halos 
around galaxies and galaxy clusters 
is now highlighted in the dark matter cosmology. 
\citet[]{Navarro97} have performed
cosmological N-body simulations
based on cold dark matter models 
and show
that the equilibrium density profiles of dark matter halos
have self-similar profiles.
The profile (hereafter, NFW profile)
is thought to be the final relaxed state  
of self-gravitating collisionless particles 
and thereby called a universal profile.
The NFW profile has a central cusp diverging as $\rho(r) \propto r^{-1}$.
Recent simulations with higher resolution 
have confirmed existence of a universal profile 
but show that 
the central density profile gets steeper and 
is better represented by $\rho(r) \propto r^{-1.5}$
\citep[]{Fukushige97, Moore00}
although slight variation of the central density profiles
from halo to halo is reported \citep[]{Jing00}.

Rotation curve measurements for dwarf galaxies 
\citep[e.g.][]{Moore94, Dalcanton00} have shown
that central density profiles are flat and
a density profile with a central cusp 
is inconsistent with the mass distribution of dwarf galaxies. 
To solve the discrepancy between the theoretical prediction 
and the observations for dwarf galaxies, 
\citet []{Spergel00} have proposed that the dark matter 
has a finite cross-section to allow 
strong self-interactions between the dark matter particles themselves.
This type of dark matter is called self-interacting dark matter
(hereafter, SIDM).
The introduction of the self-interaction may reduce 
the central mass concentration found in the universal profile 
obtained by pure collisionless dark matter simulations. 
Therefore, the precise measurement of 
central density profiles of dark matter halos is 
one of the key questions 
to solve the nature of the dark matter.

Both observational and theoretical studies have been done 
to constrain the SIDM cross-section.
Higher resolution measurements of rotation curves of dwarf galaxies 
\citep[e.g.][]{Bosch00}
have rejected the central profile of $r^{-1.5}$  
although 
the data cannot discriminate whether the central density profiles
have constant density cores or $r^{-1}$ cuspy profiles. 
Miralda-Escude (2000) has shown that 
ellipticity of the dark halo core of the lensing cluster
\objectname[]{MS 2137.3$-$2353} 
is rather high using the strong lensing data.
He has proposed to use the measured ellipticity to give 
an upper limit on the SIDM cross-section. 
\citet[]{Yoshida00} have performed N-body simulations
of cluster formation with various values of SIDM cross-sections. 
They show that 
a few collisions per particle per Hubble time at the halo center
can substantially affect the central density profile.
Therefore, precise measurements of central density profiles 
can provide a strong constraint on the SIDM cross-section.

We examine how a number ratio of radial arcs to tangential arcs
depends on the central density profile of galaxy clusters.
A radial arc is a gravitationally lensed image whose major axis lies 
in the radial direction of a cluster-lens potential. 
A tangential arc is a gravitationally lensed image whose major axis lies 
in the tangential direction of a cluster-lens potential.
We define `radial arc statistics' 
as a statistical average of number ratios of radial arcs to tangential arcs 
found in a certain cluster sample.
Recently, \citet[]{Wyithe00} have proposed an arc statistics
for probing the central density profile of galaxy clusters.
In their arc statistics,  the absolute frequency of finding arcs
appeared in a cluster must be used.
However, the absolute frequency of finding arcs 
in a certain cluster sample
is known to be sensitive to 
yet uncertain parameters e.g.\  evolution of background galaxies and
details of mass distribution of galaxy clusters
\citep[]{Hattori97, Hamana97, Molikawa99}.
No model has succeeded to reproduce the observed high frequency of finding arcs
yet.
As shown in this paper, 
the proposed radial arc statistics 
is free from 
the systematic errors coming from above uncertainties.
Throughout this paper, a cosmological model of
 $({\Omega_{\rm m}}_0, {\Omega_{\rm \Lambda}}_0, H_0) =
(0.3, 0.7, 100h\;{\rm km/s})$
is assumed.
\section{Radial arc statistics}
The density profile we use is
\begin{equation}
  \rho(r)
  = \frac{\rho_s}{\tilde{r}^a (1+\tilde{r})^{3-a}};\;\;\;\;\; 
  \tilde{r} \equiv \frac{r}{r_s},
\end{equation}
where $r_s$ is the scale radius,
$\rho_s$ is the critical density times 
the (dimensionless) characteristic over-density, 
and adopted values of $a$ are $0.5, 1.0, 1.5,$ and  $2.0$.
A projected mass $m_a(x)$ can be written as
\begin{equation}
 m_a(x) = 4 \pi r_s^3 \rho_s \int_0^\infty I_a(x,z) dz,
\end{equation}
where $\tilde{r}^2 = x^2 + z^2$ ($z$ is the line-of-sight direction.) and
\begin{equation}
I_a(x,z) \equiv \int_0^x
\frac{x^\prime dx^\prime}
{\left(\sqrt{{x^\prime}^2+z^2}\right)^a
 \left(1+\sqrt{{x^\prime}^2+z^2}\right)^{3-a}}.
\end{equation}
The lens equation reads
\begin{equation}
  y = x - b \; \widehat{\alpha}_a(x),
\end{equation}
where $y \equiv \beta/\theta_s$ is a source position, 
$x = \theta/\theta_s$ is an image position,
$b \equiv 
16 \pi G\rho_s r_s D_{\rm OL}D_{\rm LS}/c^2D_{\rm OS}
=4\rho_s r_s/\Sigma_{\rm cr}$ is called lens parameter
and $\theta_s \equiv r_s/D_{\rm OL}$.
$D_{\rm OL}$ is the angular distance from the observer to the lens,
$D_{\rm LS}$ is that from the lens to the source galaxy,
$D_{\rm OS}$ is that from the observer to the source galaxy, and
$\Sigma_{\rm cr}$ is the critical surface mass density.
For example, 
$b=1.1$ corresponds to a cluster at redshift of $0.38$
with a virial temperature $k_{\rm B}T_{\rm vir} = 8$ keV and  
a source redshift of $1.0$ \citep[e.g.][]{Eke98}.

An expected number ratio of radial arcs to tangential arcs
is calculated by calculating a ratio of 
a cross-section forming radial arcs to that forming tangential arcs.
It is assumed that source galaxies are circular
and their size is infinitesimal,
i.e.\  lensing magnification matrix 
is same everywhere in the source 
and is represented by that at the source center. 
A quantity $\lambda_{\rm r} \equiv (dy/dx)^{-1}$ represents
stretching-contracting factor 
in the radial direction of a lens potential.
A quantity $\lambda_{\rm t} \equiv (y/x)^{-1}$ represents
stretching-contracting factor 
in the tangential direction of a lens potential.
Hereafter, we use the word `radial arc' for an image
which satisfies $|\lambda_{\rm r}| > |\lambda_{\rm t}|$ and 
`tangential arc' for an image
which satisfies $|\lambda_{\rm t}| > |\lambda_{\rm r}|$.
The axis ratios of a radial arc $R(x)$ and a tangential arc $T(x)$ are
given by
\begin{equation}
 R(x) = \left|\frac{\lambda_{\rm r}(x)}{\lambda_{\rm t}(x)}\right|,\;\;
 T(x) = \left|\frac{\lambda_{\rm t}(x)}{\lambda_{\rm r}(x)}\right|,
\end{equation}
respectively.
The radial and tangential cross-sections are calculated 
by solving the following inequalities:
\begin{equation}
 \epsilon_{\rm th} < R(x) \leq \infty,\;\;\;
 \epsilon_{\rm th} < T(x) \leq \infty,
 \label{eqn:futoushiki}
\end{equation}
where $\epsilon_{\rm th}$ is the threshold axis ratio.
We calculate 
ratios of radial cross-sections to tangential cross-sections
for $a=0.5,1.0,1.5,2.0$.
The NFW profile corresponds to $a=1$.

In Figure 1, 
we show the expected number ratios of radial arcs 
whose axis ratios are larger than 
the threshold value $\epsilon_{\rm th}$, 
to tangential arcs which have the same threshold axis ratio, 
against the threshold axis ratio $\epsilon_{\rm th}$.
Calculations were done for  $b=0.5, 1.0, 1.5, 2.0,$ and $2.5$ 
which cover almost all practical combinations of
cluster redshifts and their masses, and source redshifts.

Figure 1 shows that the number ratios drastically vary with $a$ values
and variation due to variation of $b$ values is much smaller.
This indicates that 
the number ratios of radial arcs to tangential arcs 
in a certain cluster sample
is a powerful probe  
for the central density profile of galaxy clusters. 
Biases in selection of a cluster sample 
may be reflected in biases of cluster masses and their redshifts,
and the deepness of the observations which is equivalent to source redshifts. 
Changing values of these parameters only results in change of $b$ values. 
Therefore, 
insensitivity of the radial arc statistics to $b$ values shows that
the selection biases of a cluster sample less affects 
the constraint from the radial arc statistics on the central density profile. 
A dramatic increase of the radial arc number
with increasing of $a$ from $1.5$ to $2$ 
found in Figure 1 can be understood as follows.
When the central density profile is gentler than $r^{-2}$, 
a lens could have radial caustics
where the stretching factor in the radial direction become infinite, 
and radial arcs are produced in two different ways
\citep[]{Hattori99}.
When the central density profile is gentler than $r^{-1.5}$, 
radial arcs are produced 
only when source galaxies touch the radial caustics.
On the other hand,
when the central density profile is steeper than $r^{-1.5}$, 
the width of the source image along the tangential direction
is contracted.
Therefore, radial arcs can be produced without 
significant stretching in the radial direction.
As a result, the number of radial arcs are dramatically 
increased when the central density profile is steeper than $r^{-1.5}$.
Since intrinsic source size of $L_*$ galaxies
at a redshift of $\sim 1$ 
which are the dominant sources of observed arcs, 
is $\sim 1$ arcsec, 
high resolution instruments with the
Full-Width-at-Half-Maximum of better than $0.2-0.1$ arcsec., 
like {\sl Hubble Space Telescope (HST)}, {\sl Subaru} etc., 
is required to make a reliable sample
for the radial arc statistics.  
\section{Comparison with observations}
In TABLE 1, we list clusters 
for which existence of radial arcs were reported.  
In most of the cases, the radial arcs are thin and small.
Special attention is necessary to find a radial arc 
even on the {\sl HST} images. 
The listed clusters in Table 1 could be 
only a set for which existence or non-existence of radial arcs
have been securely checked. 
This sample is, of course, not uniform 
and is highly biased by the interests of the observer for each cluster. 
However, as mentioned in previous section, 
our radial arc statistics is little affected 
by the selection biases. 
We could safely use this sample for the comparison with
theoretical results.
The hyphenated arcs in Table 1 are folded or separated but nearly folded images
and, therefore, each `hyphenated arcs' is counted as a single arc 
because those (nearly) folded images are produced from a single source.
The length of these arcs is calculated 
as a half (in the case of merging of two images)
or one third (in the case of merging of three images)  
of the total length. 
Since all of the reported radial arcs have axis ratio larger than $4$,  
the threshold value can be set to $4$.
Tangential arcs with axis ratios 
larger than $4$ in the same clusters are also listed in TABLE 1.

TABLE 1 shows that 
the observational number ratio of radial arcs to tangential arcs 
is $0.7$ on average. 
This indicates $a = 1.0 \sim 1.5$.  
The central density profile with $a>1.5$ 
over-produces radial arcs and can be rejected.
On the other hand, 
the central density profile with $a<1.0$ predicts 
too few radial arcs compared with the observed results and
also can be rejected.
Therefore, the radial arc statistics 
supports that the central density profile of
cluster is in the range of $r^{-1\sim-1.5}$ and
dark matter is collisionless 
at least on the cluster scale.
\section{Discussion}
We have shown that 
the radial arc statistics which is defined as 
a number ratio of radial arcs to tangential arcs 
appeared in a certain sample of clusters,  
strongly depends on the central density profile, 
and is almost independent of other cluster properties and source redshifts.
Therefore,
our radial arc statistics which uses the number ratio can provide 
a robust test to constrain the central density profile of galaxy clusters.
A tentative comparison with observations shows that
the cluster central density profile is as steep as 
$r^{-1\sim-1.5}$. 
This result indicates that the dark matter should be 
collisionless at least in cluster scale.
Therefore, even  if the dark matter has a finite collision cross-section 
of the self-interaction $\sigma_{\rm coll}$,
the mean free time $t_{\rm free}$ of a dark matter particle (mass $m$)
should be longer than a cluster age $t_{\rm age}$.
The mean free time $t_{\rm free}$ can be estimated as 
$t_{\rm free} \simeq (\rho_s/m \sigma_{\rm coll}V_{\rm cl})^{-1}$ 
where $V_{\rm cl}\simeq 2000(T_{\rm vir}/8\;{\rm keV})^{1/2}$ km/s
is a cluster 3D velocity dispersion,
assuming that a characteristic density of the dark matter in clusters
is represented by $\rho_s$.
Assuming a cluster redshift of $0.38$,
its virial temperature of $8$ keV,
and a cluster age of $6.6h^{-1}$ Gyr,
the upper limit is obtained as
$\sigma_{\rm coll}/m < 0.11h^{-1}\;{\rm cm^2/g}$.
According to the numerical simulations \citep{Yoshida00},
this upper limit allows weak self-interaction of the dark matter
which produces a finite core in cluster mass distribution.   
Therefore, the above mentioned upper limit on $\sigma_{\rm coll}/m$
should be taken as conservative limit.
As shown by \citet[]{Yoshida00},
the collision rate expected from this small cross-section 
in the dwarf galaxies is too small   
to reproduce
core radii of dwarf galaxies 
inferred from rotation curve measurements.
Therefore, 
the existence of cores in dwarf galaxies 
is still  mystery.

Although we have assumed infinitesimal source size, 
this assumption is valid only if a variation of the 
lensing amplification within a source is negligible.
In Figure 2, we show the width of the region on the source plane
which satisfies the condition of Eq.(6) for a infinitesimal source.
The width is  scaled by $D_{\rm OS}$,
for each threshold axis ratio
using a typical value of $b=1.0$ and $r_s = 0.24\;h^{-1}$ Mpc.
Since size of $L_*$ galaxy is $\sim 1$ arcsec   
at a redshift of $\sim 1$ which is the dominant source for arcs, 
Figure 2 shows that the infinitesimal source approximation could provide 
reasonable estimations for the frequency of the tangential arcs
but  this assumption could not be good approximation
for estimating the frequency of radial arcs
for $a<2.0$.
The quantitative studies of  
the effect of the finite source extents
as well as the lens-clusters ellipticity and
irregularity in mass distribution,
will be done in the forthcoming paper.

The cluster sample used for comparison with theoretical
prediction should be improved. 
It should be mentioned, for example,  that
all the clusters listed in TABLE 1 are XD clusters.
An XD cluster is a cluster which has a bright elliptical galaxy
at the center of the X-ray emission. 
Since radial arcs are expected to appear in the very central region
of a galaxy cluster
and all the radial arcs listed in TABLE 1 are found near cDs,
it must be checked whether a 
local central density increment by the mass of the cD galaxy
plays an important role on forming radial arcs.  
Therefore much more number of clusters, including non-XD clusters,
should be included in the radial arc statistics sample
by future high resolution observation.
\acknowledgements
This research was partially supported 
by the Grants-in-Aid of 
the Ministry of Education, Science, Sports, and Culture of Japan 
(No. 11440060).
\clearpage
\begin{deluxetable}{llll}
\tablecaption{The cluster sample and observational number ratios}
\tablehead{
 \colhead{Cluster\tablenotemark{\dag}} & \colhead{Radial arc} 
 &\colhead{Tangential arc} & \colhead{Number ratio} 
}
\startdata
 \objectname[]{A 370}   & R1-2      & A0,A1-2,B2-3  & 0.3\\
 \objectname[]{AC 114}  & A4-5      & B,D           & 0.5 \\
 \objectname[]{MS 0440} & A16, A17  & A2-3, A8-9    & 1.0 \\
 \objectname[]{MS 2137} & AR        & A0            & 1.0 \\
\enddata
\tablenotetext{\dag}{Reference: A370: \citet[]{Bezecourt99}, 
 AC114: \citet[]{Natarayan98}, 
 MS0440: \citet[]{Gioia98}, 
 MS2137: \citet[]{Hammer97}}
\end{deluxetable}
\figcaption[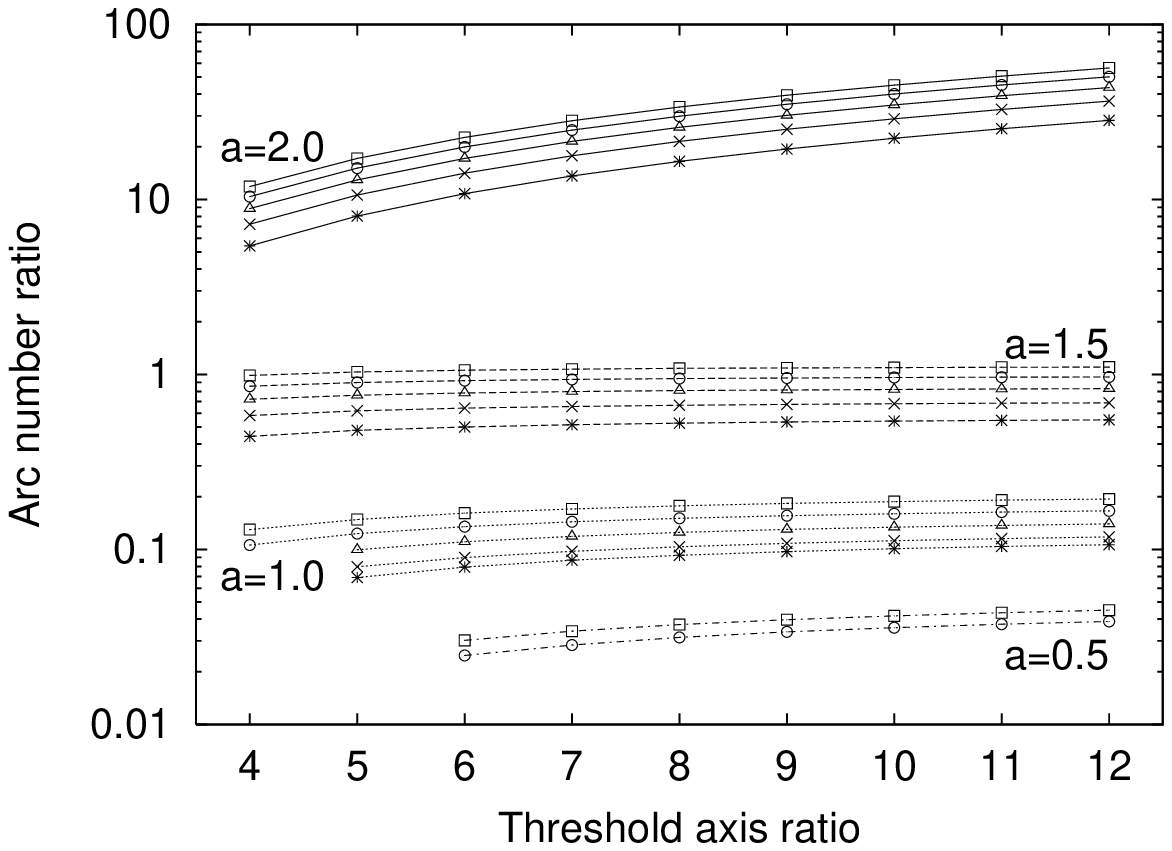]
{Ratios of radial arcs to tangential arcs 
 against threshold axis ratios.
 The uppermost five solid lines are for $a=2.0$.
 The following five dashed lines are for $a=1.5$.
 The following five dotted lines are for $a=1.0$.
 The rest two dot-dashed lines are for $a=0.5$.
 Symbols are square: $b=2.5$, circle: $b=2.0$, triangle: $b=1.5$,
 cross: $b=1.0$, and asterisk: $b=0.5$.
 If and only if $a<1$, the central surface density is finite.
 If $a<1$, $b$ must satisfy the condition $b/2(1-a)(2-a) > 1$
 to have radial arcs \citep[e.g.][]{Subramanian86}.
 The radial and tangential cross-sections are partly merge 
 where plot points are lost.
}
\figcaption[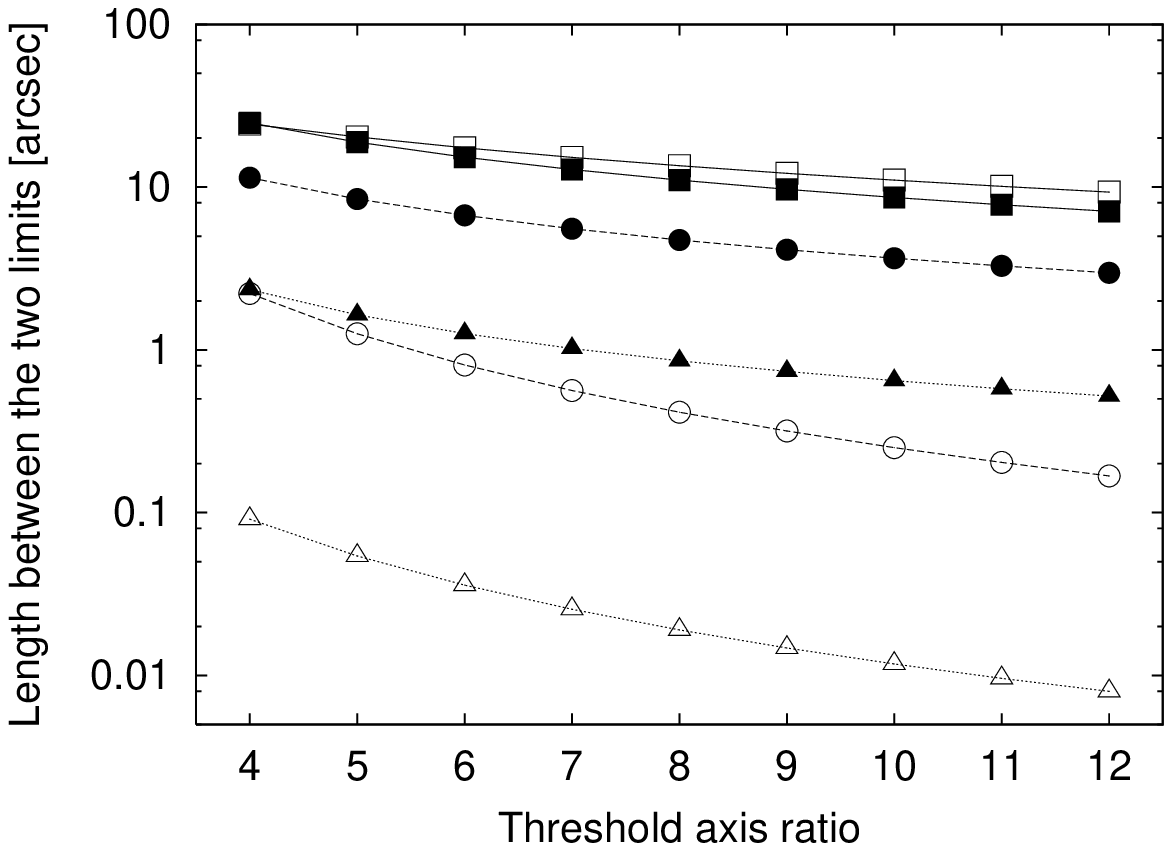]
{Angular lengths between the lower and upper limits 
 of equation (\ref{eqn:futoushiki}) on the source plane 
 against threshold axis ratios for $b=1.0$.
 The solid lines with boxes are for $a=2.0$, 
 the dashed lines with circles are for $a=1.5$,
 and the dotted line with triangles are for $a=1.0$.
 The non-filled symbols represent angular lengths between the limits for 
 $R(x)$ in equation (\ref{eqn:futoushiki}),
 and the filled symbols are for $T(x)$ in the same equation.
 Typical source size is $\sim 1$ arcsec  at a redshift of $\sim 1$. 
}
\end{document}